\documentclass{kapproc} 
\usepackage{graphicx}

\kluwerbib

\begin{document}

\articletitle[Photometric properties of LPVs 
in the LMC]{Photometric properties of long-period variables
in the Large \\ 
Magellanic Cloud}


\author{Sachiyo Noda}
\affil{
National Astronomical Observatory, 
Osawa 2-21-1, Mitaka, Tokyo 181-8588  Japan\\}
\email{sachi.t.noda@nao.ac.jp} 

\author{Mine Takeuti}
\affil{Astronomical Institute, 
Tohoku University, 
Aoba, Sendai 980-8578  Japan\\}
\email{takeuti@astr.tohoku.ac.jp}

\begin{abstract}
Approximately four thousand light curves of red variable stars
in the Large Magellanic Cloud (LMC) were selected from the
2.3-year duration MOA database by a period analysis using
the Phase Dispersion Minimization method.
Their optical features (amplitudes, periodicities,
position in CMD) were investigated.
Stars with large amplitudes and high periodicities were
distributed on the only one strip amongst multiple structure
on the LMC period-luminosity relation.
In the CMD, the five strips were located in the order of 
the period. The stars with characterized light curves were 
also discussed. 
\end{abstract}

\begin{keywords}
Red variables, period-luminosity relation, massive photometry
\end{keywords}

\section{Introduction}
The multiple and complicated period-luminosity relation for red 
variables
in the LMC had been discovered using the microlensing database
(Wood, Alcock, Allsman et al. 1999, Wood 2000).
Although the Mira sequence (Feast, Glass, Whitelock, and 
Catchpole 1989, Hughes \& Wood 1990)
have been remarked as a distance indicator,
such multiplicity is fatal for use as a distance indicator 
because their characteristics of each strip have not been revealed 
actually. The MOA (Abe, Allen, Banks et al. 1997; 
Hearnshaw, Bond, Rattenbury et al. 2000) database of the
LMC obtained by large-scale photometry is quite appropriate to study
the above problem.
Not only related with the multiplicity of the period-luminosity relation,
the photometric properties must be studied carefully to reveal 
the nature of AGB variables. Some interesting results of the 
study of the MOA database is presented. 

\section{The MOA project}
The MOA is the microlensing research project, and the
collaboration of about 30 astronomers from Japan and New Zealand.
The observational targets are mainly the LMC and Galactic bulge.
We observe every photometric night at the Mt. John University
Observatory in the center of the South Island of New Zealand 
using small telescope (61-cm diameter) and large CCD (three chips of
$2K\times 2K$ pixels) chips.
There are three data series since 1996 (Table 1).
The Series~1 is a period of test-drive, while
the Series~3 is the current system.
\begin{table}
\caption[The data series of MOA database]{The data series of MOA
 database}
\begin{tabular*}{\textwidth}{@{\extracolsep{\fill}}lccc}
\sphline
       & Series~1 & Series~2 & Series~3\cr
\sphline
Period & 1996 Jan.-1996 Dec. & 1997 Jan.-1998 Jul. & 1998 Aug.- present\cr 
Optics & f/13.5              & f/6.25   & f/6.25 \cr
FOV    & $30\prime \times 30\prime$ & $1^\circ \times1^\circ$ & $0.9^\circ \times 1.38^\circ$ \cr
CCD    & $1K\times 1K$ & $1K\times 1K$ & $2K\times 4K$ \cr
\sphline 
\end{tabular*}
\end{table}
In the current Series, the 16 fields around the LMC bar are
observed every night and approximately 4.4 million sources are 
included in the Series~3 LMC catalogue.
The curves of transmission of the two color filters and the
quantum efficiency of the CCD (SITe, $2K\times 4K$)
are shown in Noda, Takeuti, Abe et al. (2002).

\section{Red variables}
%
%

\begin{figure} 
\begin{center}
\includegraphics*[width=110mm]{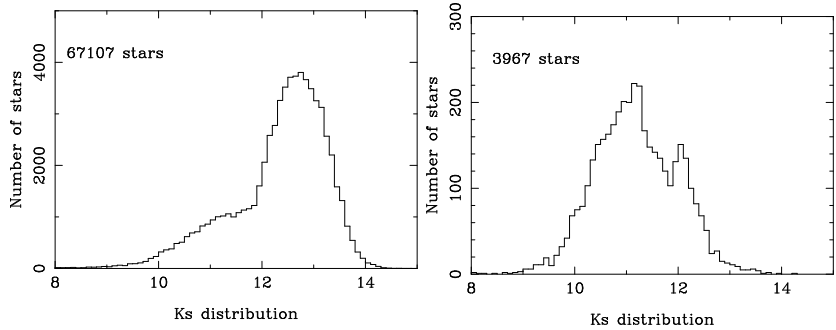}
\caption[]
{The distribution of the luminosity of non-variable (left panel)
 and variable (right panel) stars. Two clumps of variable stars are identified.}
\label{fig:fig1}
\end{center}
\end{figure}

\paragraph{Selection criteria}
The selection of regular variables from the MOA database was 
carried out as the following. 
In the first criterion (Level~1), 313,706 stars of the MOA 
database identified with only one DENIS source
(Deep Near-Infrared Southern Sky Survey; Epchtein, Deul, 
Derriere, et al. 1999) were selected. 
In the Level~2, `non-photometric' data points were eliminated,
and light curves of too small light variation were also removed 
in the Level~3.
In the Level~4 to 8, careful period analysis (PDMM) were performed 
with necessary eye-estimate to exclude inappropriate samples, and
finally, 4,858 red variables were obtained.
Amongst the selected stars at the Level~1, $K_S$ magnitudes 
(the effective wavelength $\approx$ 2.15 $\mu$m) were obtained 
for 67,107 stars. 
The histogram of $K_S$ magnitudes is indicated in the left panel
of Fig.~\ref{fig:fig1}.  

\paragraph{Two clumps of the red variables}
The $K_S$ magnitudes are tabulated for 3,967 stars among the 
finally selected variable stars. 
The histogram of $K_S$ magnitudes of these stars is shown
in the right panel of Fig.~\ref{fig:fig1}. 
The center of distribution of variable stars was $K_S \approx 11$ mag
which corresponded to a bump at the brighter side of the peak
in the left panel.
The other small peak was found around $K_S \approx$ 12 mag in the
right panel.
Because the large number of non-variables is found fainter than 
$K_S \approx 12$, it is no doubt about the existence of the 
fainter clump. The study of intrinsic properties 
of these two clumps will be important.  

\section{The period-luminosity diagram}
\begin{figure}
\begin{center}
\includegraphics*[width=77mm,angle=270]{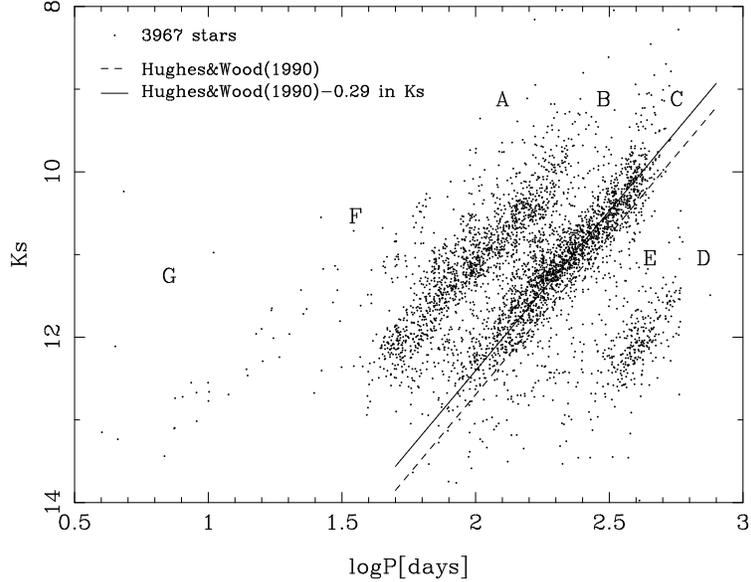}
\caption[]
{The period-luminosity diagram for 3.967 variables (see text).}
\label{fig:pl_single}
\end{center}
\end{figure}

\paragraph{Multiplicity}
The $K_S$ magnitudes as a function of $\log P$ for our 3,967 
samples is presented in Fig.~\ref{fig:pl_single}.
The dashed line is the $\log P$-$K$ relation for the oxygen-rich
Mira sequence by the previous study (Hughes \& Wood 1990), while
the solid line is the same relation but shifted upwards by 0.29 mag.
Some strips were found.
In order to investigate the structure, the vertical separation 
of $K_S$ magnitude from the PL relation by Hughes \& Wood (1990) were 
examined, and six strips (groups~$A\sim F$) were defined. 
The stars labeled with $G$ are ignored because the periods are 
very short. 
Stars in the group~$F$ were likely long period Cepheids.
3,564 stars were included in the groups~$A \sim E$.
Note, the group~$C$ was the densest strip identified with the 
classical Mira sequence.

\begin{figure}
\begin{center}
\includegraphics*[width=80mm,angle=270]{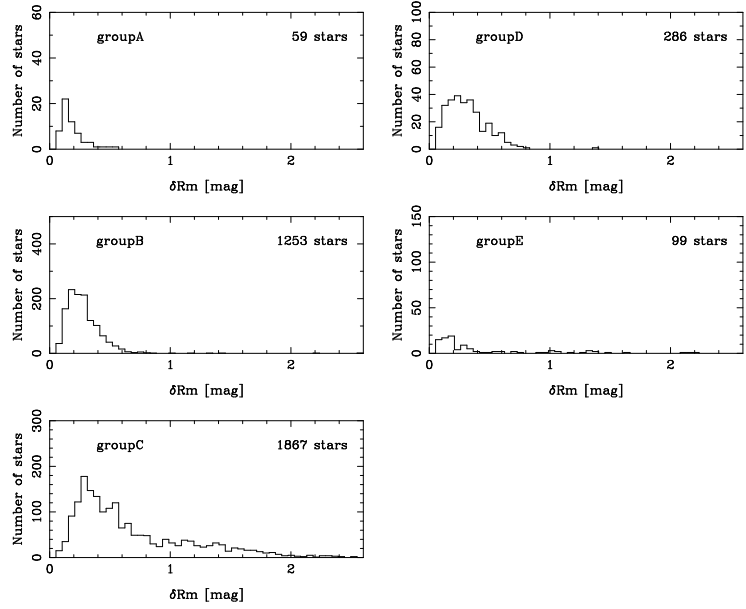}
\caption{The histogram of amplitude, $\delta R_m$, for the five 
strips.}
\label{fig:amp_hist}
\end{center}
\end{figure}

\paragraph{The distribution of amplitude}
In Fig.~\ref{fig:amp_hist}, the amplitude 
histograms of the group~$A\sim E$ are presented.
The amplitude $\delta R_m$ was defined as the difference of the 
magnitudes of the brightest and the faintest bins. 
In this figure, stars of large amplitude were mainly belonged 
the group~$C$.
For example, 95\% of stars whose amplitude were
larger than 0.9 were the member of group~$C$.

\begin{figure}
\begin{center}
\includegraphics*[width=80mm,angle=270]{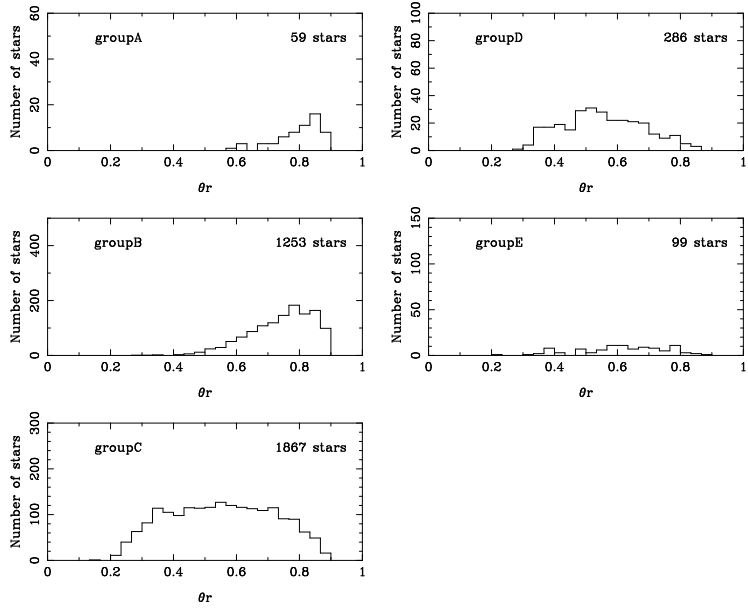}
\caption {The histogram of periodicities, $\theta _r$, for the 
five strips.}
\label{fig:theta_hist}
\end{center}
\end{figure}

\paragraph{Regularity of the period}
The histograms for the periodicity, $\theta _r$, histogram are presented
in Fig.~\ref{fig:theta_hist}.
$\theta _r$ ($0 < \theta _r < 1$) is the relative parameter
which is defined in the PDMM and indicates regularity
of the light variation. Small $\theta _r$ 
indicates high regularity, while large $\theta _r$ 
indicates almost random variation.
In this figure, the distribution of $\theta _r$ for group~$C$ 
showed it to be almost constant for $0.2 < \theta _r < 0.9$.
It is obvious that stars with small $\theta _r$, for example,  
88\% of stars with $\theta _r < 0.4$ were the member of group~$C$.

\paragraph{The most remarkable sequence}
Amongst stars which were satisfied the both condition of
$\delta R_m > 1$ and $\theta _r < 0.4$, 96\% were the group~$C$ 
component.
This means the absolute magnitudes of long-period variables will
be estimated properly when the sufficient number of
light curves over several cycles are obtained.
The luminosity of variables of 
large amplitude an d high regularity must be estimated by using
the period-luminosity relation of group $C$.

\section{The ($\langle R_m\rangle - K_S, K_S$) diagram}
\begin{figure}
\begin{center}
\includegraphics*[width=90mm,angle=0]{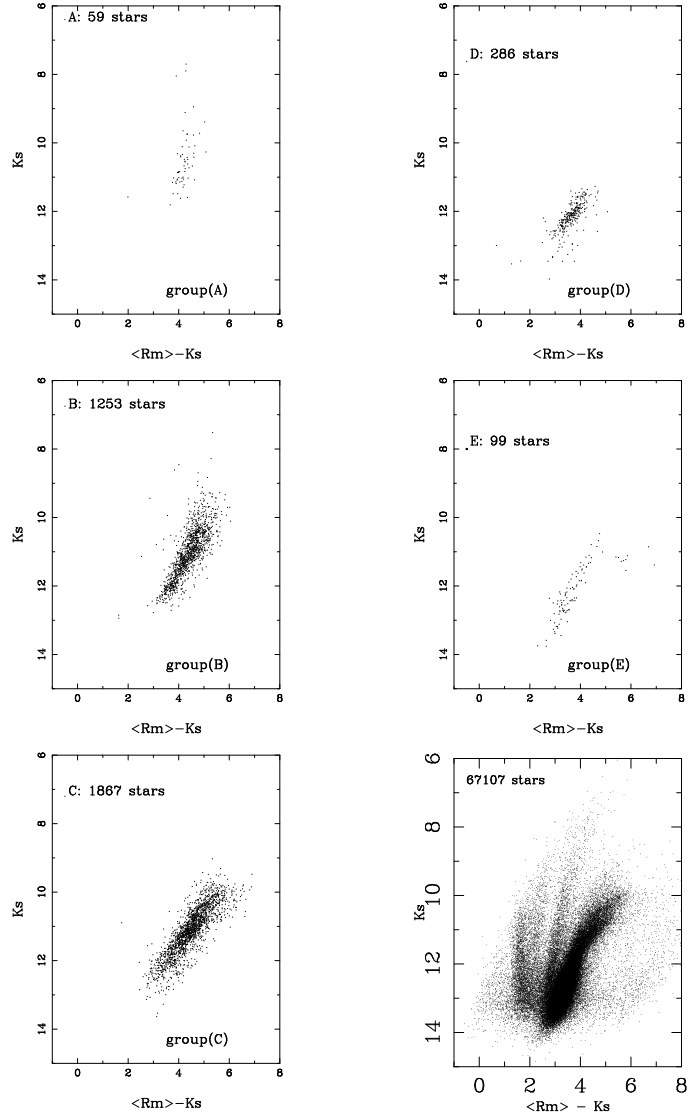}
\caption{The ($\langle R_m\rangle$ - $K_S, K_S$) diagram. The 
bottom-right panel shows the distribution of 67,107 stars which
were identified by DENIS and detected at least in $K_S$ magnitude.}
\label{fig:CMD}
\end{center}
\end{figure}

The color magnitude diagrams, ($\langle R_m\rangle - K_S, K_S$),
for each group are shown in Fig.~\ref{fig:CMD}.
$\langle R_m\rangle$ indicates the mean magnitude in the MOA red
light curve.
In this figure,
the group name and the number of sample are presented in the 
each panel.  The bottom-right panel indicates the distribution of 67,107
red stars which were identified by DENIS and detected at least in $K_S$ 
magnitude.
Not only variables but also non-variables are plotted in this panel.
As reported in the preliminary study by Noda et al. (2002), the 
variables were located in almost same domain on the CMD. 

Because the present samples are very rich in the number, precise
comparison of the properties of each group will be interesting. 
The mean $K_S$ magnitudes of groups $A$ - $C$ correspond to the 
bright clump beside those of groups $D$ and $E$ are identical 
with the faint clump. The mean positions of each group on the CMD 
differ group to group too. Group $A$ is brighter and bluer than 
group $B$, and the latter is also brighter and bluer than group 
$C$. 
The examined regression lines showed the slope of the lines 
for group~$A, B$ and $C$ were arranged in the period order 
(The slope of group$~A$ is steeper than group $B$ and so on). 
In the groups of the faint clump, group $D$ is the fainter and 
redder one, and the slope of group $E$ is slightly steeper than group
$D$. 

\section{Other interesting results}

\paragraph{RV Tauri-like feature}
The light curves of RV Tauri stars are characterized by the 
alternative deep and shallow light minima. We have found that 
13.5 \% of the variable star samples show this typical RV Tauri 
type light curve. It is interesting that the majority of these stars 
(94 \% !) belong to group $C$. Because the luminosity of these stars 
are typical to the Mira stars, it is clear that these stars are not
the RV Tauri type.
It will be mentioned that the RV Tauri-like light curve alone is
not the characterized property for the RV Tauri stars.

\paragraph{Eclipsing variables}
In the process of selection, we have found many stars showing 
the light curve typical to the eclipsing binary system. 
In 348 such stars, 159 stars are located on the same position 
as group $D$, and 103 stars are as group $C$, 
on the ($\log P$, $K_S$) diagram. 
The existence of many eclipsing binaries 
at the same position as group $D$ on the ($\log P$, $K_S$) 
diagram is a new enigma about the AGB stars. 
It should be noted that group $D$ of our paper is established
after excluding the eclipsing binaries. The nature of group $D$
stars will be studied without the connection of the eclipsing binaries.

\paragraph{Period transition}
Difference in the periods was found between the stars tabulated 
in Hughes \& Wood (1990) and the present results. 
Such a difference was also found between the result of Noda et 
al. (2002) and the present one. It is found that the period 
transition of Galactic semi-regulars in the solar neighborhood
was also common in the LMC red variables. 
Together with the RV Tauri-like light curves, these feature will
be evidence for multi-mode behavior of the AGB variables. 


\section{Conclusion}

Almost all of large amplitude 
or highly periodic stars in the LMC were the member of group~$C$
which is the most crowded and nearest strip to the classical Mira 
sequence. For example, the 94\% stars of large amplitude
($\delta R_m > 1$) or 88\% stars of inferior periodic ($\theta _r$ $< 0.4$)
were the group~$C$ component.
If we require the both conditions,  more than 90\% were the 
member of group~$C$ strip.
It is possible to utilize this type of sequence as a distance
indicator, even if multiple relations are found in extra-galactic
systems.

In the ($\langle R_m\rangle - K_S$, $K_S$) diagram, the stars of
the five sequences show slightly different features in the order of 
period. Because all of the variables were found in a limited 
domain of the CMD, the intrinsic excitation mechanism will be 
the same, but the difference in the pulsation mode is suggested.

\end{document}